\documentclass[usenatbib]{mn2e}
\usepackage{graphicx}

\begin{document}
\topmargin-1cm

\newcommand{\apj}{ApJ}
\newcommand{\apjl}{ApJL}
\newcommand{\apjs}{ApJS}
\newcommand{\mnras}{MNRAS}
\newcommand{\aap}{AAP}
\newcommand{\prd}{PRD}
\newcommand{\aj}{AJ}
\newcommand{\pasp}{PASP}
\newcommand{\araa}{ARA\&A}

%%%%%%%%%%%%%%%%%%%%%%%%%%%%%%%%%%%%%%%%%%

\newcommand{\be}{\begin{equation}}
\newcommand{\ee}{\end{equation}}
\newcommand{\bea}{\begin{eqnarray}}
\newcommand{\eea}{\end{eqnarray}}
\newcommand{\Msun}{M_{\odot}}
\newcommand{\A}{\AA\,}
\newcommand\lesssim{\mbox{$^{<}\hspace{-0.24cm}_{\sim}$}}
%%%%%%%%%%%%%%%%%%%%%%%%%%%%%%%%%%%%%%%%%%

\title{Dust attenuation in the restframe ultraviolet: constraints from
  star--forming galaxies at $\lowercase{z}\sim1$}

\author[C. Conroy] {Charlie Conroy\\ Department of Astrophysical
  Sciences, Princeton University, Princeton, NJ 08544, USA}

\date{\today}
\maketitle

\begin{abstract}

  A novel technique is employed for estimating attenuation curves in
  galaxies where only photometry and spectroscopic redshifts are
  available.  This technique provides a powerful measure of particular
  extinction features such as the UV bump at $2175$\AA, which has been
  observed in environments ranging from the Milky Way to
  high--redshift star--forming galaxies.  Knowledge of the typical
  strength of the UV bump as a function of environment and redshift is
  crucial for converting restframe UV flux into star formation rates.
  The UV bump will impart a unique signature as it moves through
  various filters due to redshifting; its presence can therefore be
  disentangled from other stellar population effects.  The utility of
  this technique is demonstrated with a large sample of galaxies drawn
  from the DEEP2 Galaxy Redshift Survey.  The observed $B-R$ color of
  star--forming galaxies at $0.6<z<1.4$ disfavors the presence of a UV
  bump as strong as observed in the Milky Way, and instead favors
  restframe UV ($1800$\A$<\lambda<3000$\A) attenuation curves similar
  to the Milky Way without a UV bump, a power--law with index
  $\delta=-0.7$, or a form advocated by Calzetti and collaborators.
  Stronger constraints on the strength of the UV bump in galaxies can
  be achieved if independent constraints on the $V$-band optical depth
  are available.

\end{abstract}

%-----------------------------------------------------------------

\section{Introduction}
\label{s:intro}

Estimating the stellar masses, ages, and star formation rates of large
samples of galaxies has become common thanks both to large homogeneous
spectroscopic and photometric surveys and to increasingly accurate and
sophisticated stellar population synthesis (SPS) models.  These models
rely on stellar evolution calculations, stellar spectral libraries, an
initial stellar mass function, and accurate dust attenuation models.
Unfortunately, each of these ingredients carry uncertainties that are
large enough to significantly impact the derived physical properties
\citep[see e.g.][and references therein]{Conroy09a}.

Accurate modeling of attenuation by dust is particularly challenging.
It is common practice in SPS to constrain the dust opacity in the
visible portion of the spectrum, and then adopt an attenuation curve
to infer the dust opacity at both shorter and longer wavelengths.
Recall that attenuation differs from extinction in that the latter
describes the loss of photons along a given line of sight due to
either scattering or absorption.  Attenuation, in contrast, refers to
the net loss of photons, and for simple geometries is equivalent to
the loss of photons due to true absorption.  In general an attenuation
curve includes the complex radiative transfer effects due to the
star--dust geometry as well.  It is therefore the concept of
attenuation that is most relevant to modeling the integrated light
from galaxies.

Common assumptions for the wavelength dependence of attenuation
include a power--law: $\tau\propto\lambda^{\delta}$ with
$\delta=-0.7$, \citep{Charlot00}, a parameterization advocated by
Calzetti and collaborators \citep{Calzetti94, Calzetti00}, the
extinction curves of the Milky Way and Magellanic Clouds
\citep{Cardelli89, Fitzpatrick99}, or attenuation curves derived via
the combination of either the Milky Way or Magellanic Cloud extinction
curves with a variety of star--dust geometries \citep{Witt00,
  Gordon01}.  An attenuation curve must be assumed --- rather than
self--consistently applied --- because the dependence of dust
properties on quantities such as metallicity, local UV radiation
intensity, and local and large--scale star--dust geometry are not
understood either observationally or theoretically with the precision
required for SPS models \citep{Draine03}.

Direct constraints on the wavelength--dependence of dust obscuration
come principally from two techniques.  The first is formally a probe
of dust extinction, and assumes that the intrinsic spectrum of a
source, such as a star, quasar, supernova, or gamma ray burst (GRB),
is known, and then the ratio between the observed and intrinsic
spectrum provides a probe of the wavelength--dependent extinction.
This method is commonly used to measure extinction in nearby galaxies
where individual stars can be resolved.  It has also been applied to
high--redshift galaxies that by chance alignment are back-lit by GRBs
\citep[e.g.][]{Ardis09}, to multiply lensed quasars
\citep[e.g.][]{Motta02}, and supernovae \citep[e.g.][]{Riess96}.

The second technique, which formally probes dust attenuation, relies
on the identification of classes of galaxies with similar physical
properties (such as star formation rates and metallicities) that
differ only in dust content.  Comparing dusty to dust--free galaxies
in the same class then provides a measure of the attenuation curve for
the dusty galaxies within the class.  This technique was pioneered by
Calzetti and collaborators to measure the attenuation curve of
actively star--forming galaxies at low redshift \citep{Calzetti94}.
Variants of this technique have been applied more recently by
\citet{Johnson07b, Johnson07a} to photometry of a large sample of low
redshift galaxies, and \citet{Noll09} who analyzed spectra of
star--forming galaxies at $z\sim2$.

The most striking feature of the extinction curves in the Milky Way
and Large Magellanic Cloud is the strong, broad absorption feature at
$2175$\AA, the `UV bump' \citep{Stecher65}.  The carrier of this
absorption feature has not been positively identified, though
polycyclic aromatic hydrocarbons (PAHs) are a leading candidate
\citep{Weingartner01, Draine03}.  Curiously, there is little evidence
of a UV bump through most sightlines in the SMC.  Multiply lensed
quasars and gamma-ray bursts, when used as probes of foreground
galaxies, show varied results regarding the presence of the UV bump
\citep{Motta02, Vijh03, Wang04, Mediavilla05, York06, Ardis09}.

Searching for evidence of a UV bump in the integrated light from
galaxies is significantly complicated by the fact that the effects of
scattering and geometry will tend to alter the underlying extinction
curve \citep{Witt92}.  In simple geometries, radiative transfer
effects will make the effect of the UV bump stronger (assuming that
the UV bump is not due to scattering), while in more complex
configurations the strength of the UV bump can be significantly
weakened \citep{Witt00}.  Regardless of the geometrical configuration,
if the UV bump is present in the underlying extinction curve, it's
presence should be detectable in the integrated light from galaxies.
Observations of the strength of the UV bump in external galaxies
therefore constrains a mixture of the underlying extinction curve and
the geometrical configuration of stars and dust in the galaxy.

The UV bump appears to be absent from the net attenuation curves of
local starburst galaxies \citep{Calzetti94, Calzetti00}, leading
\citet{Witt00} to suggest that in such galaxies the underlying dust
extinction curve lacks a UV bump.  A detailed analysis of M51 also
finds little evidence for a UV bump within individual HII regions
\citep{Calzetti05}.  Curiously, spectroscopy of a sample of
star--forming galaxies at $z\sim2$ shows evidence for the UV bump
\citep{Noll05, Noll09}, indicating both that the underlying extinction
curve must have a UV bump and that the star--dust geometry does not
significantly dilute its strength.

The strength of the UV bump encodes unique information regarding the
formation and destruction of dust grains and is therefore a useful
probe of the interstellar medium (ISM) where the UV photons are being
absorbed \citep{Draine03}.  Explanations for the observed variance in
the strength of the UV bump include metallicity dependence, the
effects of complex star--dust geometry, and varying physical
conditions of the ISM \citep{Gordon03}.

The observed variation in strength of the UV bump has interest beyond
the ISM.  UV flux is routinely used as a proxy for star formation in
galaxies.  The observed UV flux is corrected for dust attenuation via
the adoption of an attenuation curve which, in most cases, do not
include the UV bump.  However, if this additional absorption feature
is present in galaxies, omission of it in the adopted attenuation
curve will result in a systematic underestimation of star formation
rates.  For example, the NUV, $u$, $B$, and $R$ band filters at
$z\approx0.0,0.6,1.0,$ and $2.0$, respectively, are sensitive to
the presence of this feature.

The present work is dedicated to exploring the observational evidence
for a strong absorption feature at 2175\A in a large sample of
star--forming galaxies at $z\sim1$.  Where necessary, a flat
$\Lambda$CDM cosmology is assumed with the following parameters
$(\Omega_m,\Omega_\Lambda,h) = (0.24,0.76,0.72)$.  All magnitudes are
in the $AB$ system \citep{Oke83}, and stellar masses assume a
\citet{Chabrier03} initial mass function (IMF).

\section{Technique}
\label{s:tech}

Imagine identifying the same, or statistically similar galaxies at a
variety of redshifts.  The flux through a fixed bandpass filter will
then trace out the average spectral energy distribution (SED) of the
sample because of redshifting.  In effect, one is measuring the
average SED for this set of galaxies convolved with the filter.  As
the UV bump moves into and out of a given filter, it will impart
unique changes that, with the aide of models, can be separated from
variations in the underlying continuum.  This basic idea was exploited
by \citet{Vijh03} to conclude that average star--forming galaxies at
$z\sim2$ do not show evidence for a strong UV bump.  A similar
technique has also been employed by \citet{Assef08} to construct
average SED templates from broadband photometry of galaxies at $0<z<1$.

The challenging aspect of this approach is the requirement that one
identifies the same, or similar galaxies at multiple epochs.  This
requirement is, however, relatively easy to satisfy in the restframe
UV.  As demonstrated in the following section, for typical
star--forming galaxies the restframe UV is sensitive principly to dust
attenuation and is not sensitive to other physical parameters such as
total stellar mass or metallicity.

This technique is probing the spectrum of an {\it average}
star--forming galaxy and therefore avoids several of the standard
degeneracies between dust attenuation and star formation rate (SFR)
when fitting individual galaxies.  For example, the SFH of an average
galaxy must be smooth, and one therefore cannot appeal to recent
bursts of star formation in the interpretation of the restframe UV.
In addition, this technique becomes conceptually cleaner at higher
redshifts because at higher redshifts the redshift range over which
the UV bump would move into and out of a given filter will correspond
to a smaller change in lookback time.  One may thus more confidently
gather a statistically similar sample of galaxies over the requisite
redshift range.

\section{Models}
\label{s:bump}

\begin{figure}
\begin{center}
\resizebox{3.3in}{!}{\includegraphics{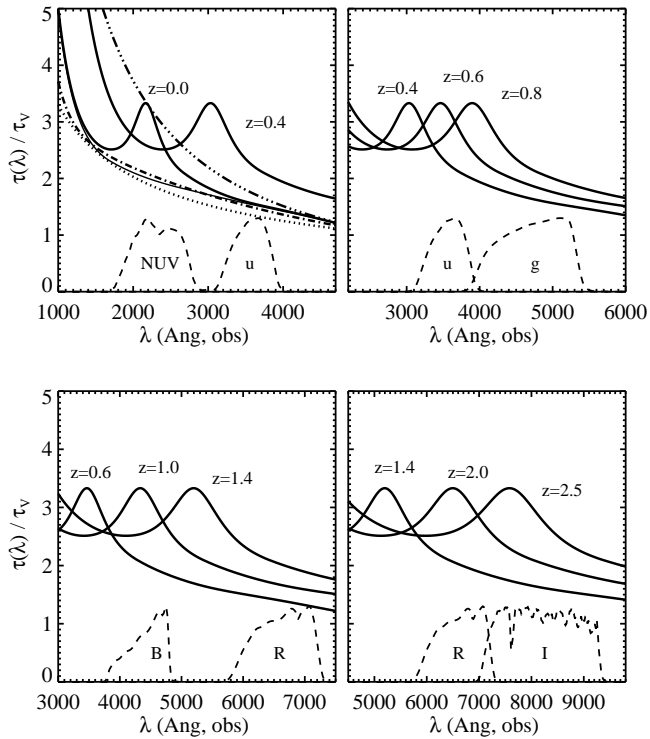}}
\end{center}
\caption{Average Milky Way extinction curve ({\it solid lines}),
  normalized to the $V-$band, as would be observed at various
  redshifts.  Transmission curves (with arbitrary normalization; {\it
    dashed lines}) are included for common bandpass filters that would
  be sensitive to the presence of a UV bump.  A power--law attenuation
  curve with index $\delta=-0.7$ ({\it dotted line}), $\delta=-1.3$
  ({\it dot--dot--dot--dashed line}), Calzetti attenuation ({\it
    dot--dashed line}), and a Milky Way curve without a UV bump ({\it
    thin solid line}) are included in the upper left panel.  Note the
  different $x-$axis scales in each panel.}
\label{fig:ext}
\end{figure}

This section describes the SPS model, including the prescription for
attenuation by interstellar dust, used to generate SEDs of mock
galaxies.  The relation between attenuation curves and observed colors
is explored, and a demonstration of the proposed technique is
undertaken.

The SPS treatment closely follows that of \citet{Conroy09a}, to which
the reader is referred for details.  The model includes all relevant
phases of stellar evolution for metallicities in the range
$10^{-4}<Z<0.030$, for ages $10^{6.6}<t<10^{10.2} $ yrs, and for
initial masses $0.10\leq M\leq67 \,\Msun$ \citep{Marigo07a, Marigo08}.
The IMF of \citet{Chabrier03} is adopted.

The fiducial dust model closely follows the two component model
proposed by \citet{Charlot00}, although a wider variety of attenuation
curves will be considered herein.  Stars younger than $10^7$ years are
subject to attenuation associated with their birth cloud.  In
addition, all stars experience attenuation due to diffuse, cirrus
dust.  These two sources of attenuation are characterized by $\tau_1$
and $\tau_2$, respectively, where $\tau$ is the optical depth at
$\lambda=5500$\AA.  This two component model has strong observational
motivation not only from direct observations of young stars embedded
in molecular clouds but also from integrated spectra of star--forming
galaxies, where the opacity measured in balmer emission lines is a
factor of $\sim2$ larger than the opacity measured from the stellar
continuum \citep{Calzetti94}.  \citet{Charlot00} favor $\tau_1=1.0$
and $\tau_2=0.3$ based on a sample of low redshift star--forming
galaxies.  Unless stated otherwise, these will be the parameter
choices used herein.

An attenuation curve must be adopted to extrapolate the optical depth
at 5500\A to both shorter and longer wavelengths.  A variety of
commonly used curves are considered for the cirrus dust, including a
power--law with index $\delta$, the average extinction curve measured
for the Milky Way, both with and without the UV bump at 2175\AA, and
an intermediate case where the UV bump strength is one half of the
nominal Milky Way value.

A second dust model is also considered.  In this model the attenuation
curve of Calzetti et al. is adopted and is applied to all starlight
independent of stellar age.  When employing this model, the single
opacity characterizing the dust optical depth is equal to $2.5\tau_2$.
The numerical coefficient was chosen to achieve a close correspondence
in restframe UV colors between the two models.  Our conclusions are
not sensitive to this particular value.

In summary, when the Calzetti attenuation curve is used, all starlight
is attenuation equally, independent of stellar age.  Where all other
attenuation/extinction curves are mentioned, the two--component dust
model of \citet{Charlot00} is employed.  In this model the dust around
young ($t<10^7$ yrs) stars is assumed to have a power--law attenuation
curve with $\delta=-0.7$ \citep{Charlot00}, except at the end of
$\S$\ref{s:res} where other curves are considered.  It is the
attenuation curve of the cirrus, diffuse dust that is varied between a
power--law and Milky Way--type curves.

Figure \ref{fig:ext} shows the Milky Way extinction curve as seen at
various redshifts.  The figure also shows the location of various
bandpass filters, and, in the top left panel, the attenuation curve of
Calzetti et al. and power--law attenuation curves with index
$\delta=-0.7$ and $\delta=-1.3$, for comparison.  The purpose of this
figure is to highlight at which epochs particular filters are
sensitive to the presence of the UV bump.  For example, at $z=1.0$ the
UV bump would be redshifted into the $B-$band, and so by considering
the $B-$band flux of galaxies at $0.6<z<1.4$ one might hope to measure
the strength of the UV bump.

\begin{figure}
\begin{center}
\resizebox{3.5in}{!}{\includegraphics{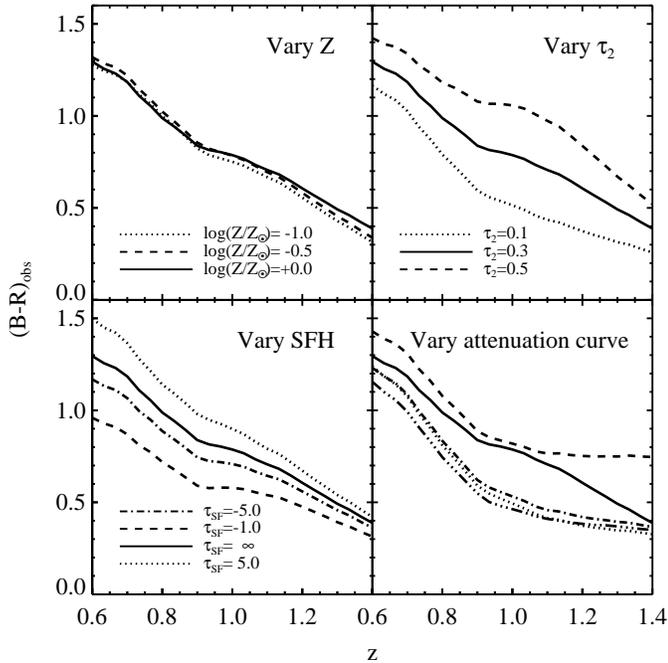}}
\end{center}
\caption{Observed--frame $B-R$ color as a function of redshift.  In
  each panel one parameter is varied about the default values of
  $Z=Z_{\odot}$, constant SFH ($\tau_{\rm SF}=\infty$), $\tau_2=0.3$,
  and a Milky Way extinction curve with a UV bump.  The lower right
  panel shows the effect of varying the attenuation curve between a
  power--law with an index of $-1.3$ ({\it dashed line}) and $-0.7$
  ({\it dot--dashed line}), a Calzetti attenuation curve ({\it
    dot--dot---dot---dashed line}), and Milky Way extinction curves
  with ({\it solid line}) and without ({\it dotted line}) a UV bump.}
\label{fig:varyuv}
\end{figure}

The power of the proposed technique to constrain the average restframe
ultraviolet attenuation curve is demonstrated with mock galaxies.
Each mock galaxy is assumed to be composed of stars of a single
metallicity and with a SFR characterized by a simple exponential:
SFR$\propto e^{-t/\tau_{\rm SF}}$ from $t=0$ to the age of the
universe at the redshift of the mock galaxy.  Starlight is attenuated
by dust according to the model described above.  For reference, for a
constant SFH the young and old components contribute equal flux at
$\lambda=2500$\AA (restframe) in the absence of dust.

The evolution of observed--frame $B-R$ colors with redshift is shown
for mock galaxies in Figure \ref{fig:varyuv}.  In this figure the
sensitivity of the $B-R$ color evolution to various parameters is
explored.  It is clear that the color evolution is entirely
insensitive to metallicity, and is only sensitive to the star
formation history (SFH) for pathological values of $\tau_{\rm SF}$.
Color evolution is most sensitive to the normalization of the
attenuation curve, $\tau_2$ and the attenuation curve itself.  While
not shown, the color evolution is also insensitive to $\tau_1$ for
$0.5<\tau_1<1.5$.

\section{Data}
\label{s:data}

Galaxies are drawn from the DEEP2 Galaxy Redshift Survey
\citep{Davis03}, which has gathered optical spectra for $\sim40,000$
galaxies, primarily in the redshift range $0.7<z<1.4$.  Target
galaxies were selected using $BRI$ imaging from the CFHT telescope
down to a limiting magnitude of $R=24.1$ \citep{Coil04b}.  In three of
the four fields observed colors are used to exclude objects likely to
have $z<0.7$; the sampling rate at $z<0.7$ is thus $1/4$ the rate at
higher redshift.  Redshift errors are $\sim30$ km s$^{-1}$ as
determined from repeated observations.  Details of the DEEP2
observations, catalog construction, and data reduction can be found in
\citet{Davis03}, \citet{Coil04b}, and \citet{Davis05}.  Restframe
$U-B$ colors and absolute $B$-band magnitudes, $M_B$, have been
derived as described in \citet{Willmer06}.  Stellar masses for a
subset of DEEP2 galaxies were derived by \citet{Bundy06} utilizing
ancillary $K-$band data.  With these masses an empirically derived
relation between rest--frame $UBV$ colors and stellar mass was
obtained in order to assign stellar masses to all DEEP2 galaxies
(C.N.A. Willmer, private communication).

Star formation rates, metallicities, and $V-$band dust opacities are
not readily available for all DEEP2 galaxies.  Galaxies thus cannot be
binned in these quantities as a function of redshift.  Instead, two
simple cuts are made to ensure that similar galaxies are selected
across the redshift range.  A cut on restframe $U-B$ color is made to
exclude red sequence galaxies: $U-B<-0.032(M_B+21.63)+1.03$
\citep{Willmer06}.  A second cut is made on stellar mass, requiring
$10.0<{\rm log}(M/ \Msun)<10.5$.  Since stellar mass is correlated
with SFR at $z\sim1$ \citep{Noeske07a}, this cut corresponds, at least
approximately, to a cut on SFR.  Applying these cuts to the data
leaves $4,203$ galaxies in the redshift range $0.6<z<1.4$.  These cuts
result in a sample that is volume limited until $z=1.2$.  Beyond this
redshift, the apparent $R-$band cut preferentially selects against
intrinsically red objects.  Fortunately, none of the conclusions
herein rely on the data points at $z>1.2$.  Moreover, the results
presented in the following section are robust to the subsample of data
used.  For example, the conclusions are unchanged if all galaxies are
included in the analysis.

\section{Results}
\label{s:res}

\begin{figure*}
\begin{center}
\resizebox{6.0in}{!}{\includegraphics{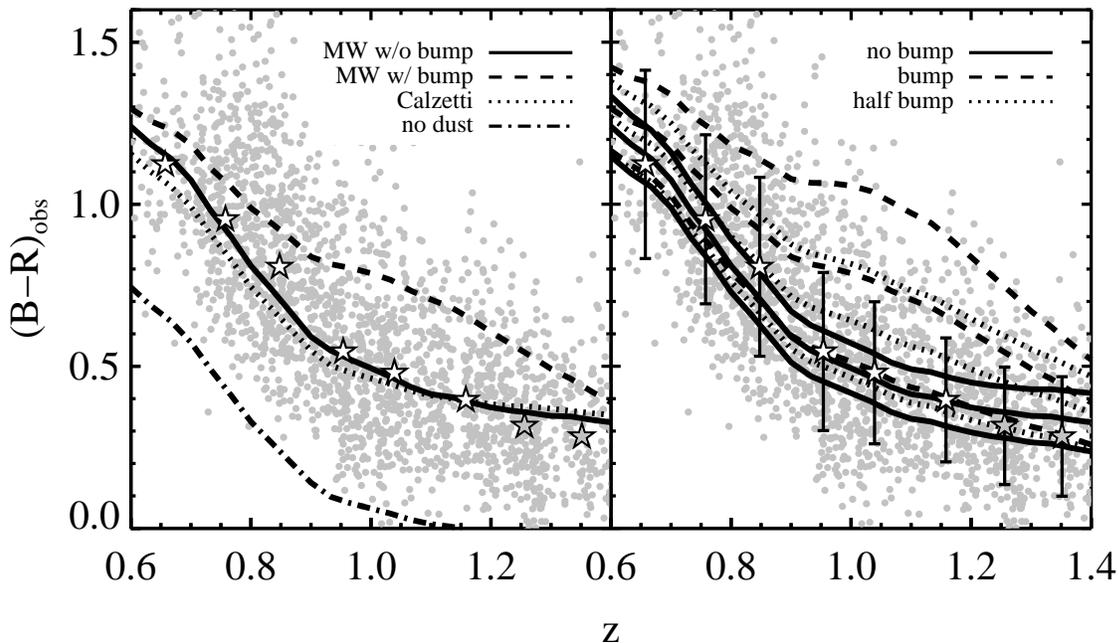}}
\end{center}
\vspace{0.3cm}
\caption{Observed--frame $B-R$ color as a function of redshift.  Data
  are from the DEEP2 Survey ({\it solid grey circles}).  Model
  predictions ({\it lines}) are compared to the average trend in the
  data ({\it stars}; stars are white at $z<1.2$ where the data are
  volume limited).  Error bars denote $1\sigma$ scatter in the data.
  {\it Left panel:} Data are compared to model predictions for the
  standard set of assumptions ($\tau_2=0.3$, constant SFH,
  $Z=Z_{\odot}$), with three dust attenuation curves: Milky Way
  extinction with ({\it dashed line}) and without ({\it solid line})
  the UV bump, and Calzetti attenuation ({\it dotted line}); a
  dust--free model is included for reference ({\it dot--dashed line}).
  {\it Right panel:} Data are compared to model predictions using
  Milky Way extinction curves with ({\it dashed lines}) and without
  ({\it solid lines}) the UV bump, and with a UV bump intermediate in
  strength ({\it dotted lines}), for three values of the dust opacity:
  $\tau_2=0.1,0.3,0.5$ ({\it bottom, middle, and top lines},
  respectively).  The Milky Way extinction curve without a UV bump
  provides the best description of the data over a wide range in
  $\tau_2$, while the extinction curve with a UV bump only describe
  the trends in the data for unreasonably low values of $\tau_2$ (see
  $\S$\ref{s:res} for details).}
\label{fig:brdat}
\end{figure*}

The primary result of this article is contained in Figure
\ref{fig:brdat}.  In this figure the variation in observed $B-R$ color
with redshift for $4,203$ DEEP2 galaxies is compared to the expected
color evolution for a galaxy with solar metallicity and constant star
formation rate.  In the left panel the data are compared to expected
colors for the default dust model parameter $\tau_2=0.3$, for three
different attenuation curves: Milky Way extinction both with and
without a UV bump, and Calzetti attenuation.  In this panel a
dust--free model is included for comparison.

In the right panel attention is focused on models with Milky Way
extinction curves both with and without the UV bump, and a curve with
a UV bump strength one half of the full value, for
$\tau_2=0.1,0.3,0.5$.  It is clear from this figure that the presence
of the UV bump with strength comparable to that observed in the Milky
Way is strongly disfavored by the data, unless the typical galaxy at
$z\sim1$ has a $V-$band opacity of $\tau_2\leq0.1$.  The latter
possibility is highly unlikely given that typical $V-$band opacities
for star--forming $\sim L^\ast$ galaxies are rarely lower than 0.2
both in the local universe and at higher redshift \citep[for a review,
see][]{Calzetti01}.  If an attenuation curve with a Milky Way--type UV
bump is to be retained, then the distribution of $B-R$ colors at fixed
redshift implies not only that the average $\tau_2$ be $\leq0.1$ but
that the distribution of $\tau_2$ values must be very strongly peaked
about $\tau_2=0.1$.  This must be so in order for there to be so few
data points above $B-R=0.7$ at $z=1.0$, for example.  Such a scenario
seems implausible.

The figure also demonstrates that a Milky Way extinction curve without
a bump can fit the data with a variety of dust optical depths.  This
should be regarded as a positive feature since galaxies are observed
to span a range in optical depths \citep{Wang96, Calzetti01}.
Moreover, our results can easily accommodate the picture that dust
opacity scales with the blue luminosity of galaxies, as is observed in
the local Universe \citep[][]{Wang96}.  Blue luminosities are on
average higher at $z\sim1$ compared to $z\sim0$, and so while local
galaxies are characterized by typical values of $\tau_2=0.3$
\citep{Charlot00}, galaxies at higher redshift may well have higher
$\tau_2$.  Dust opacities of $\tau_2=0.5$ or higher are well within
the observational constraints and therefore moderate evolution in the
dust opacity to higher redshift can be tolerated so long as a UV bump
is not prominent in the dust attenuation curve.

The models with a UV bump strength that is one half the nominal Milky
Way value produce observed $B-R$ colors that are approximately
intermediate between the full bump and no bump models.  In this case
an optical depth of $\tau_2=0.3$ is much less discrepant with the
data, although the range in model colors between $0.1<\tau_2<0.5$ is
still rather large given the likely range of optical depths in
galaxies at $z\sim1$.  Nonetheless, it is clear that more refined
constraints on the strength of the UV bump must await independent
estimates of the restframe $V-$band optical depth in these galaxies.

It is apparent from the left panel of Figure \ref{fig:brdat} that the
model using the Calzetti attenuation curve also provides a good
description of the observed trend.  This should not be surprising in
light of Figure 1, where it is clear that the Calzetti attenuation
curve is similar to the Milky Way curve without a UV bump over the
relevant wavelength range.

Comparison of the lower right panel of Figure \ref{fig:varyuv} with
Figure \ref{fig:brdat} reveals several additional results.  First, the
Milky Way extinction curve without the UV bump is very similar to a
power--law dust attenuation curve with index $\delta=-0.7$, at least
over the restframe wavelengths probed in the figure.  Therefore,
power--law attenuation curves with $\delta=-0.7$ also agree well with
the data.  On the other hand, a power--law attenuation curve with
$\delta=-1.3$ produces colors much redder than observed for
$\tau_2=0.3$.  Models with $\delta=-1.3$ and $\tau_2=0.1$ produce
acceptable agreement with the data, but they are then subject to the
same objections as raised above for the Milky Way extinction curve
with the UV bump.

These results are not sensitive to the dust attenuation curve adopted
for young stars.  Recall that the fiducial dust model associates
additional dust obscuration around stars with ages $t<10^7$ years.  A
power--law attenuation curve was adopted for these young stars.  If
instead a Milky Way extinction curve with a UV bump is adopted for all
stars, including young ones, then the disagreement between the models
and data is somewhat more severe.  Furthermore, varying the $V-$band
optical depth around young stars from $\tau_1=0.5$ to $\tau_1=1.5$
does not change these results.  Varying the boundary between young and
old stars from $10^{6.5}$ to $10^{7.5}$ years, or allowing for a
fraction of `naked' young stars that have escaped their birth clouds,
shifts the model predictions by a constant amount in $B-R$ and
therefore cannot help to reconcile the model predictions including a
UV bump with the data.  Only when $\tau_1$ approaches 0.0 do the model
predictions become sensitive to the treatment of attenuation around
young stars.  However, such low values for $\tau_1$ are unphysical
given not only the measured extinction toward O stars in the Milky Way
but also the observed ratios of balmer emission lines in local
star--forming galaxies \citep{Calzetti94}.

It is important to remember that $BR$ photometry at these epochs is
probing restframe $1800$\A$<\lambda<3000$\AA.  The results herein
concerning the attenuation curve therefore only apply to that
wavelength range.  It would thus be unwise to extrapolate these
results to longer or shorter wavelengths.

\section{Conclusion}
\label{s:disc}

If the UV bump at restframe 2175\A were a generic feature of
extragalactic attenuation curves, then the observed--frame $B-R$
colors of galaxies should redden substantially as the UV bump
redshifts into the $B-$band at $z=1$, and then rapidly become bluer as
this feature redshifts out of the band by $z=1.4$.  Comparison of
simple models to the observed $B-R$ colors of galaxies over the
redshift range $0.6<z<1.4$ reveals that the UV bump with a strength
comparable to that observed in the Milky Way is not a ubiquitous
feature in the observed SEDs of star--forming galaxies at $z\sim1$.

A more general lesson to be drawn here is that dust attenuation curves
need not be assumed, but instead can be directly constrained by the
data, even when the data consist only of broadband photometry.  It
would be fruitful to repeat this exercise at both lower and higher
redshifts, where the necessary data already exists.  More refined
constraints on the strength of the UV bump can be expected if
independent estimates of the $V-$band dust opacity are available.
Constraints on dust attenuation in the restframe ultraviolet for large
samples of galaxies at a variety of epochs will not only afford a more
robust transformation between UV flux and SFR, but may also shed light
on dust properties and the relative distribution of stars and dust
over a range of environments and epochs.

\section*{Acknowledgments}

It is a pleasure to thank Bruce Draine for numerous conversations
related to this work, Jeffrey Newman for clarifying various data
related issues and catching several errors in an earlier draft, and
Christopher Willmer for providing his derived data products catalog.
The referee, Daniela Calzetti, is thanked for comments that
significantly improved the clarity and content of the manuscript.
Funding for the DEEP2 survey has been provided by NSF grants
AST95-09298, AST-0071048, AST-0071198, AST-0507428, and AST-0507483 as
well as NASA LTSA grant NNG04GC89G.  The data presented herein were
obtained at the W. M. Keck Observatory, which is operated as a
scientific partnership among the California Institute of Technology,
the University of California and NASA. The Observatory was made
possible by the generous financial support of the W. M. Keck
Foundation.  This work made extensive use of the NASA Astrophysics
Data System and of the {\tt astro-ph} preprint archive at {\tt
  arXiv.org}.

\bibliographystyle{mn2e}
%\bibliography{../master_refs}

\end{document}